\documentclass[sigconf]{acmart}

\usepackage{float,graphicx}
\usepackage{subcaption}
\usepackage{booktabs} 
\usepackage{multirow}
\usepackage{enumitem,acmart-taps}
\usepackage{tikz}
\usepackage{natbib,xcolor,framed}
\usepackage{tikz}

\definecolor{Aquamarine}{RGB}{75,160,200}

\colorlet{TFFrameColor}{black}
\colorlet{TFTitleColor}{white}

\newenvironment{takeawaybox}[1][Takeaway]{%
  \begin{titled-frame}{#1}
}{%
  \end{titled-frame}
}

\newcommand{\titlebox}[2]{%
  \par\medskip
  \noindent
  \begin{tikzpicture}
    \node[
      draw=gray!60,
      line width=0.8pt,
      fill=gray!10,
      inner sep=0pt,
      text width=\linewidth
    ] (box) {%
      \begin{tikzpicture}
        \node[
          fill=black!80,
          text=white,
          font=\bfseries,
          inner xsep=-4pt,
          inner ysep=5pt,
          anchor=west,
          minimum width=\linewidth,
          align=left
        ] {\hspace*{-12pt}\vbox{#1}\hspace*{-20pt}};
      \end{tikzpicture}
      \par
      \vspace{6pt}
      \hspace{8pt}%
      \parbox{\dimexpr\linewidth-16pt\relax}{#2}
      \vspace{8pt}
    };
  \end{tikzpicture}
  \par\medskip
}

\newcommand{\tool}{\emph{Privacy Cards}}

\keywords{Older adults; longitudinal study; smart speaker; voice assistant; Alexa; privacy; ethics; cards}

\copyrightyear{2026}
\acmYear{2026}
\setcopyright{cc}
\setcctype{by}
\acmConference[CHI '26]{Proceedings of the 2026 CHI Conference on Human Factors in Computing Systems}{April 13--17, 2026}{Barcelona, Spain}
\acmBooktitle{Proceedings of the 2026 CHI Conference on Human Factors in Computing Systems (CHI '26), April 13--17, 2026, Barcelona, Spain}
\acmPrice{}
\acmDOI{10.1145/3772318.3791893}
\acmISBN{979-8-4007-2278-3/2026/04}

\begin{document}

\title[Privacy Cards for Surfacing Mental Models and Exploring Privacy Concerns]{Privacy Cards for Surfacing Mental Models and Exploring Privacy Concerns: A Case Study of Voice-First Ambient Interfaces \\with Older Adults}

\author{Andrea Cuadra}
\orcid{0000-0002-1845-0240}
\affiliation{
  \institution{Olin College}
  \city{Needham}
  \state{MA}
  \country{USA}}

\author{Samar Sabie}
\orcid{0000-0001-9904-6081}
\affiliation{\institution{University of Toronto}
\city{Mississauga}
\state{Ontario}
\country{Canada}}

\author{Yan Shvartzshnaider}
\orcid{0000-0001-9904-6081}
\affiliation{%
  \institution{York University}
    \city{Toronto}
  \state{ON}
  \country{Canada}}

\author{Deborah Estrin}
\affiliation{
\institution{Cornell Tech} 
  \city{New York}
  \state{NY}
\country{USA}}

\begin{abstract}

We investigate the ethical and privacy implications of voice-first ambient interfaces (VFAIs) for aging in place through an in-depth engagement with five older adults. Our participants were in the process of becoming experienced VFAI users, and had used a VFAI-based design probe for health data reporting. We create and iteratively refine an interview protocol using \tool. We customize \tool~ by drawing on participants’ previous interviews and device usage logs. Using \tool, we conduct interviews to surface their mental models, and explore their privacy concerns. We find insufficient mental models for proper consent. For example, participants did not know who could access their data, and experienced difficulty distinguishing built-in functionality from third-party apps. Participants initially expressed little worry about VFAI-related ethical concerns, but interviews with \tool~ revealed nuanced issues, resulting in various implications for future research and design.
\end{abstract}

\ccsdesc[500]{Human-centered computing~Empirical studies in HCI}
\ccsdesc[500]{Security and privacy~Usability in security and privacy}
\ccsdesc[500]{Human-centered computing~User centered design}
\ccsdesc[500]{Human-centered computing~Participatory design}

\maketitle

\section{Introduction}

As the proportion of older adults continues to increase, so do the technological innovations that promise to support them as they age in place. Voice assistants (e.g., Amazon’s Alexa, Google’s Assistant, Apple’s Siri, ChatGPT voice), which can serve as VFAIs, may have exciting applications for decreasing isolation, motivating physical activity, providing continuity of care, calling for emergency assistance, among others. VFAIs\footnote{VFAIs utilize ambient intelligence \cite{cook2009ambient}, which employs advanced sensors and sensor networks, pervasive computing, and artificial intelligence to (passively) integrate into daily human life by making people’s surroundings flexible and adaptive. In this work, we use commercially-available smart speakers with screens to study VFAIs.} are ubiquitous and pervasive information hubs. 

However, VFAIs’ inner workings are difficult to discern, and are being increasingly deployed to unknowing users (e.g., older adults in nursing homes or senior centers). VFAIs introduce risk of harm, such as by sharing information across space, groups of people, and time in ways that can threaten privacy and generate other ethical issues. For example, any person's interactions with an Amazon Alexa\footnote{Amazon Alexa is chosen as the example here due to its extensive use and popularity.} are by default saved and recorded to the cloud on its associated Amazon account. This means that specific requests made to the VFAI, the tone used to make these requests, any background noises that were audible during the request, and all conversations that ensued during instances when the VFAI may have been activated by accident are accessible to anyone with the password to that Amazon account. A person (e.g., child, grandchild, technical helper) with this password may have access to an older adult's seemingly private interactions captured through their VFAI. Many VFAIs, despite having a single voice, rely on third-party applications from thousands of creators across various contexts, such as health and education, each capable of collecting information about how users interact with their devices from the privacy of their homes. Some of this information is collected without necessarily granting the individuals whose data is at play the ability to delete it, modify it, or know about it.

\enlargethispage{20pt}
Moreover, computational models can make far-reaching inferences about people from their data in ways that are difficult for individuals to predict \cite{nissenbaum2019contextual}. Relying on history to signal the future, the appearance of increasingly capable technologies will continue to cross, blur, and shift existing contextual boundaries, exacerbating users’ vulnerability to privacy violations. As some scholars argue, merely asking participants if they have privacy concerns is not sufficient and can be potentially misleading \cite{martin2016measuring, barkhuus2012mismeasurement}, because there is little opportunity to ground responses in participants’ lived experiences in an interactive way that is simultaneously reflective, participatory, and theory-informed. Playful and participatory methods that are more reflective and grounded in lived experiences exist, such as the cards that~\citet{Gaver1999_probes} use as design probes; however, they tend to be more generative, making them unsuitable for eliciting privacy judgments in a theory-informed manner that facilitates clarification and revision. This creates an urgent need to innovate within Human-Computer Interaction (HCI) to design alternative approaches for exploring risks and mitigating potential harms, particularly for groups that have not been well-represented in the design of these technologies. 

Responding to this need, we create \tool~by taking inspiration from the contextual integrity (CI) theory of privacy~\cite{nissenbaum2009privacy}. 
The CI theory defines privacy as the appropriate flow of information, where appropriateness is determined by established contextual information norms (or privacy norms). According to CI, privacy is potentially violated when information flows breach these norms. \tool~specifically draw on the CI framework to capture information flows using five essential parameters, allowing for a flexible and systematic way to express information flows and evaluate their alignment with existing privacy norms.

\citet{anon2023}  discuss the promise of VFAIs for aging in place, describing the stories of five older adult participants who became voice assistant users through their study and with whom they speculated about future interfaces for health and wellbeing through three interviews. The authors caution that VFAIs pose privacy risks, calling for thoughtful, careful, and systematic consideration of these issues---yet they do not provide findings on privacy itself. With IRB approval for sharing data, we use the interview recordings and device usage logs from \citet{anon2023} to inform our study with the same participants. \textbf{Our goal is to create and use \tool---a novel, card-based interview protocol---as both a methodological and exploratory tool to surface privacy perceptions and ethical challenges in older adults’ interactions with VFAIs, while also advancing inclusive and ethical design practices for emerging ambient technologies.} We chose this approach because existing options were either too complicated for our participants, or not specific enough to address our concerns of potential VFAI harms. We chose design cards because they provide great promise for encoding design knowledge \cite{aarts2020design}, utilizing a human-centered approach to regulation \cite{luger2015playing}, and broadening participation and sharing of power \cite{elsayed2023responsible, Borning2012_values} in a simple and user-friendly manner. We specifically investigate the following research questions:
\enlargethispage{20pt}
    \begin{enumerate}
        \item[\textbf{RQ1:}] \textbf{Methodological focus.} How can the \tool~ interview protocol help researchers elicit older adults’ mental models\footnote{A mental model is a user's understanding of a system \cite{carroll1988mental}.} and concerns about privacy in VFAIs?
        \item[\textbf{RQ2:}] \textbf{Ethical focus.} What ethical and privacy-related issues emerge when older adults engage with VFAIs using the \tool~ protocol? 
        \item[\textbf{RQ3:}] \textbf{Implications.} How might insights from using \tool~ inform the ethical design and study of VFAIs for older adults?
   \end{enumerate}

We make three main contributions to CHI: we 1) create \tool~ and describe their iterative design process, which explains how we adapted them to older adults based on their usage logs; 2) uncover several ethical concerns of using VFAIs for older adults, such as insufficient mental models that affect their ability to properly consent; and 3) identify implications for design and research, including using \tool~ to inform large-scale survey design. Our novel empirical findings and their associated implications for the research and design of VFAIs provide evidence of the effectiveness of \tool. This exploratory work is an important step toward designing more ethical VFAIs for older adults.

\section{Related Work}
Our work lies at the intersection of privacy ethics, CI theory, research on older adults’ use of VFAIs, data inference, and card-based design methods. While each of these areas has been studied independently, we connect them to address a central challenge: how to co-create meaning and insights with older adults about opaque data practices in everyday interactions with VFAIs using a theoretically grounded and participatory method. We argue that CI is powerful but abstract, older adults are an important yet underrepresented group of users in digital privacy studies about VFAIs, data inference makes risks invisible, and card methods can bridge this---but only if designed considering all of these intersections.


\subsection{Existing ethical concerns about digital privacy}
A key issue with digital privacy is that we do not yet have ethical, privacy-preserving solutions that adequately address the risks introduced by ubiquitous information technologies. Questions surrounding deception, stereotypes, privacy, and accountability permeate the literature \cite{strengers2020smart, lacey2019cuteness, stark2016emotional}. Moreover, there are well-documented gaps in users’ understanding of VFAIs that are already in widespread use \cite{horstmann2023alexa}, increasing the risk of privacy violations. These gaps may be intensified for groups that have been systematically excluded from the design of digital technologies, such as older adults \cite{stypinska2023silicon}. At the same time, older adults refine mental models through exploration and flexible support strategies, though the design of the technology strongly influences how accessible and effective those strategies are \cite{sharifi2025helping}.

Despite the increasing interest in exploring privacy design from an HCI perspective~\cite{kumar2024roadmap}, these issues remain largely understudied. Several attempts move us closer to better solutions, but the wickedness of the problem \cite{ritchey2013wicked} requires much more. For example, \citet{alhirabi2023parrot} implemented and evaluated a proof-of-concept prototype to facilitate the incorporation of privacy specific design features into the IoT application
from the beginning rather than retrospectively, an important step but certainly not enough. 

The information that people disclose to machines, either directly or indirectly, can be used in both good and bad ways.
For example, knowing that you may have a certain illness sooner rather than later, can result in better health outcomes. Similarly, knowing a person's interests by analyzing what content or activities they have been engaging with can help tailor future information and recommendations to those interests, making future experiences more enjoyable. However, these usage patterns can also expose a person’s vulnerabilities and fears, which can be nefariously used to many ends, such as fueling panic and division, or tricking people into paying for goods or services they do not want or need. This information can also be problematic if it is used in a harmful manner (e.g., by a predatory insurer). These sorts of assessments can also result in decisions that affect people without their knowledge, and without giving people the ability to revise errors or exercise their rights \cite{cinnamon2017social}. Information revealed to a machine can become part of a larger repository of information that is managed by entities with commercial interests, which can be misaligned with the interests of the individual disclosing that personal information \cite{zuboff2015big}. Often, the information shared with the machine is stored, and available to others \cite{Dellinger19}, likely unbeknownst to the user. 

Furthermore, some argue that the ability of technology to manipulate information, bypass long-standing privacy values, and influence behavior through social machines cannot be adequately addressed by traditional privacy protection mechanisms \cite{calo2009people}. 
It is unclear which legal regimes should govern these technologies and what consumer protection rules for them should look like \cite{hartzog2014unfair}. 

In this work, we respond to these concerns by using a design-focused approach to explore, identify, and unveil potential harms in a highly contextualized manner with older adult participants, generating rich questions for further exploring the ethics of VFAIs. 

\subsection{Contextual integrity (CI) theory of privacy}~\label{relatedwork-privacy}
Privacy concerns are not easily measurable, calling for more nuanced treatments of the notion of privacy within HCI \cite{barkhuus2012mismeasurement}. One way to achieve this is by drawing from theoretical frameworks, such as CI \cite{nissenbaum2009privacy}. CI can help researchers explore privacy norms in social contexts. Privacy norms shape our expectations for what is appropriate in a given situation---they are socially constructed and dynamic, and they may change when new technologies create new social contexts \cite{proferes2022development}. 

CI has been used in HCI for different purposes. For example, \citet{shvartzshnaider2016learning} used a CI-based vignette study~to understand privacy norms in an educational context. \citet{apthorpe2018discovering} used a similar approach to study privacy expectations of Smart Home devices. In further work, \citet{apthorpe2019evaluating} used the CI-vignette study method to understand parents' privacy expectations about the use of devices protected by the Children's Online Privacy Protection Rule. To explore the ethical dimension of big data research project,  \citet{zimmer2018addressing} developed a CI-based heuristic to help researcher think about relation of privacy, anonymity and harm. 

We chose to create \tool, because there are not many easily accessible strategies to facilitate the use of CI for developing a nuanced understanding of the \textit{``why''} for groups that have been marginalized and underrepresented in the design of digital technology. In today's world, avoiding using digital technology is increasingly difficult, making it more urgent to uncover privacy norms in different demographics and cultural contexts. Recently, \citet{kumar2024roadmap} developed a roadmap for HCI researchers on using CI in conducting qualitative privacy research. This work serves as an example of this roadmap in action towards guiding HCI and social computing researchers on how to apply CI to qualitative projects.

\subsection{Older adults' use and perception of VFAIs}
Although work examining older adults’ use and perceptions of VFAIs has increased in recent years \cite{harrington2022s, so2024they, chung2025developing, pradhan2025no, pradhan2025understanding, brewer2024intersectional, karimi2025designing, mathur2025sometimes, anon2023}, older adults remain substantially underrepresented in VFAI research. Reviews of the computing literature found that 97\% of ACM Digital Library publications did not consider age or aging in voice user interface studies \cite{stigall2019older, stigall2020systematic}. At the time these reviews were conducted (2019–2020), nine percent of the global population was over 65\footnote{\url{https://data.worldbank.org/indicator/SP.POP.65UP.TO.ZS}}
, indicating roughly a threefold underrepresentation. As a result, VFAIs do not adequately represent the needs of older adults, which may put them at higher risk of harm.

\citet{stigall2020systematic} further observed that most studies reported older adults taking more time to operate voice user interfaces and/or making more errors than younger adults, with a minority finding no age-related differences. A more recent systematic review highlights older adults' visions for future AI-based conversational systems, expressing a desire for more human-like interactions, personalization, and greater control over their information \cite{huang2025designing}. The desire for more human-like interactions and personalization are notable, because they may increase the amount of trust in the VFAIs alongside the amount of data collected. Moreover, designing for greater control over user's information requires detailed understanding of their conceptual models of the technology. Our work dives deeper into these matters while especially considering the emotional limitations and potential dangers associated with anthropomorphized conversational agents~\cite{cuadra2024illusion,devrio2025taxonomy}.

Other studies document a mismatch between media portrayals of VFAIs and older adults’ lived experiences. \citet{sin2022does} found that older adults perceived VFAIs as more primitive, limited, and difficult to set up than mass-media narratives suggest. Such discrepancies may create feelings of inadequacy and reinforce stereotype threats around older adults’ technology use. Relatedly, \citet{shandilya2022understanding} found that while older adults expressed enthusiasm for AI-enabled products, including VFAIs, they also worried about privacy intrusions and a loss of agency.

Research specifically examining older adults’ privacy concerns with VFAIs remains relatively scarce. A comparative study of adults under 65 and 65+ found that older adults reported fewer VFAI-related privacy concerns regarding consent, data security, and data protection~\cite{spangler2022privacy}, raising questions about whether they fully understood the associated risks—a theme we investigate here. Similarly, \citet{bonilla2020older} found that older adults expressed concerns about VFAI data practices but were often unaware of available privacy resources, concluding with a call to better support this population. Finally, \citet{so2024they} found that older adults had many privacy and ethical concerns about future VFAI technologies, but did not explore their mental models about how VFAIs currently work. Our work  addresses these concerns in a novel manner by examining how \tool~ can surface gaps in understanding and reveal the ethical challenges older adults face when interacting with VFAIs.

\subsection{Data inference}
 The ability to produce inferences from data elevates the need to approach this problem in theory-grounded, participatory manner. Health status information may be inferred from inconspicuous sensors embedded in our built environments~\cite{pan2019fine}, wearable technologies~\cite{fedor2023wearable, jaques2017predicting}, digitally mediated social networks~\cite{khalid2023exploiting}, keystroke patterns~\cite{vesel2020effects}, and various other not necessarily medical contexts. According to \citet{turow2021voice}, many scientists believe that a person's weight, height, age, and race, potential illnesses they may have, can also be identified from the sound of that individual's voice. Moreover, human machine interactions and information disclosures go beyond the individual, meaning that data a person discloses can inadvertently disclose information about others as well~\cite{choi2019privacy,nissenbaum2019contextual,turow2018media}. VFAIs in particular contain a large amount of information about people, such as the particular times of day when people use certain commands, users' explorations (or lack thereof) of new topics, and how their usage patterns relate to those of other users~\cite{bentley2018understanding}. 

The amount of information we can derive from inconspicuous, or seemingly innocuous data sources, highlights the importance of studying potential harms of emerging technologies, in particular for marginalized groups, such as older adults. This work illustrates a way to do so, and can serve as an initial foundation to determine how to mitigate potential harms.

\subsection{Use of card activities to inform digital design}
HCI has long concerned itself with the use of cards as a design tool, especially as a way for designers to encode and communicate design knowledge \cite{aarts2020design}. Over time, card-based approaches have expanded to support a wide range of design goals. Tools such as \textit{Envisioning Cards} \cite{friedman2012envisioning} and the \textit{IDEO Method Cards}\footnote{\url{https://www.ideo.com/journal/method-cards}}
 illustrate how cards can catalyze creativity, surface values, and prompt new perspectives. Synthesizing this landscape, \citet{hsieh2023cards} identified seven forms of design knowledge embedded in 161 card decks—from creative inspiration to values in practice—and noted that most decks emphasize early-stage ideation. Our \tool~ may support design at all stages. Card sets have also been used to support ethical and inclusive design, such as \textit{Playing the Legal Card} \cite{luger2015playing} and \textit{Responsible \& Inclusive Cards} \cite{elsayed2023responsible}, which help practitioners engage with challenging topics and democratize participation. Other HCI decks illustrate this breadth: the \textit{Tarot Cards of Tech} \cite{tarotcards} use speculative prompts to help teams anticipate unintended consequences; \textit{Judgment Call The Game} \cite{ballard2019judgment} engages players in scenario-based ethical decision-making; and the \textit{Building Utopia} toolkit~\cite{bray2022radical} supports community-led envisioning of equitable technological futures. Together, these examples highlight cards’ flexibility as reflective, ethical, and participatory design tools.

Early work such as \citet{Gaver1999_probes} introduced Cultural Probes---open-ended postcard kits intended to spark inspiration and surface unexpected ideas with older adults. While powerful for generative design, their ambiguity makes them unsuitable for surfacing or structuring the nuanced, theory-driven privacy judgments required in our work. More recent efforts explore card-based methods for eliciting such judgments. For example, \citet{berkholz2025playing} created \textit{Privacy Taboo}, a card game designed to serve as a playful breaching interview method, fostering discourse on unwritten privacy rules. When playing \textit{Privacy Taboo}, their younger adult participants articulated their information needs when consenting to fictive data requests, even when contextual cues were limited. Our \tool~ builds on these traditions by grounding elicitation in older adults’ lived experiences while also systematically leveraging a privacy framework to surface concrete information needs. We rely on cards' playful design affordances to foster engagement and make discussions of challenging topics more accessible, drawing directly from participants' VFAI usage logs and interviews. By doing so, we tie together privacy ethics, the CI theory, research on older adults’ use of VFAIs, and data inference risks towards creating a deeper understanding of ethical concerns that regard older adult use of VFAIs.

\smallskip
\noindent
\textbf{Taken together, prior work highlights a critical gap at the intersection of the domains described in this section.} CI offers a rigorous framework for reasoning about privacy, and card-based methods provide accessible, participatory tools for ethical engagement. Yet CI has, to the best of our knowledge, not been operationalized in a way that is specifically tailored to older adults' interactions with digital technologies (less so with VFAIs specifically). Meanwhile, studies of older adults’ VFAI use document privacy concerns and desires for control, but lack methods for uncovering the conceptual foundations required for a meaningful conception of privacy. The need for such a method is amplified in light of data inference practices that make information flows invisible. In this work, we use the CI to create \tool, which addresses this critical gap by building on prior work to surface older adults’ mental models and privacy concerns in real contexts.
\section{\tool~ design process}
To design \tool~ we combined user tests, senior center field observations, and consultations with experts in CI, older adults, and individuals outside computing or adjacent fields. Additionally, we incorporated information from earlier participant interviews and usage log tracking.

This study relies on other research about designing VFAIs to support aging in place that was led by the same first author, where the authors conducted a two-month field deployment of a VFAI \cite{anon2023}. During this period, they tracked participants' usage of their VFAIs and interviewed them on different topics. Their interviews covered the following topics in chronological order: 1) general use and familiarization, 2) home health, and 3) wellbeing. \citet{anon2023} shared their primary data with us with permission from all IRBs involved. In this paper, we report the findings from a privacy interview that we carried out between their home health and wellbeing interview, using familiarity from the prior engagements and device usage logs. 

We now describe our design process. First we briefly explain CI and the five parameters that it requires (Sender, Recipient, Subject, Attribute, Transmission Principle). These parameters are the core of CI, so they became the core of ~\tool{}. We describe the details of how we adapted the terminology used in CI to make it more accessible to a wider audience. For example, from talking to older adults and individuals outside computing or adjacent fields, we determined that CI uses academic language that can be difficult for many to understand, so we worked to simplify the language. We also selected values for each parameter that we considered were relevant to our participants' interactions with their VFAIs. Finally, we describe how we developed the mechanics of \tool~ and how we then digitized them.


\subsection{Contextual Integrity}\label{relatedwork-privacy} We initially studied Nissenbaum's seminal book, \textit{Privacy as Contextual Integrity} \cite{nissenbaum2009privacy}. CI posits that privacy is not maintained by keeping an information secret, adhering to a well-defined procedure, providing specific access controls, or seeking informed consent. 
CI defines privacy in terms of the appropriateness of information flows in a given context, as prescribed by governing contextual norms. Potential privacy violations occur when information flows deviate from established norms or societal expectations. As connected technologies are introduced into our homes, our information is exposed beyond the home context, connecting a typically personal indoors setting into a globally reachable and observed environment. This notion of the ``appropriateness of an information flow'' is ever more important when technologies generate information flows that users are unaware of and that may violate the privacy norms of a specific context.

To examine the privacy implications of these technologies, the CI framework 
identifies information flows that deviate from the norms and uses the CI heuristic to evaluate their contribution to the purpose, values, and function of a given context. CI analysis requires  capturing individual information flows and norms using five parameters: 
\begin{enumerate}
\item \textbf{Sender:} the actor that sends the information
\item \textbf{Recipient:} the actor that receives the information
\item \textbf{Subject:} the actor of whom the information is about
\item \textbf{Attribute:} this is the information type, in our study we call it ``data''
\item \textbf{Transmission Principle:} this states the constraints on the flow, such as sharing confidentially or with subject's consent.
\end{enumerate}

A noteworthy aspect we debated was how to categorize the VFAI. We ultimately decided to categorize it as a communication medium, like a telephone device. 


\subsection{Adapting CI to cards for older adult participants}
\begin{figure*}
\centering
  \includegraphics[width=\textwidth]{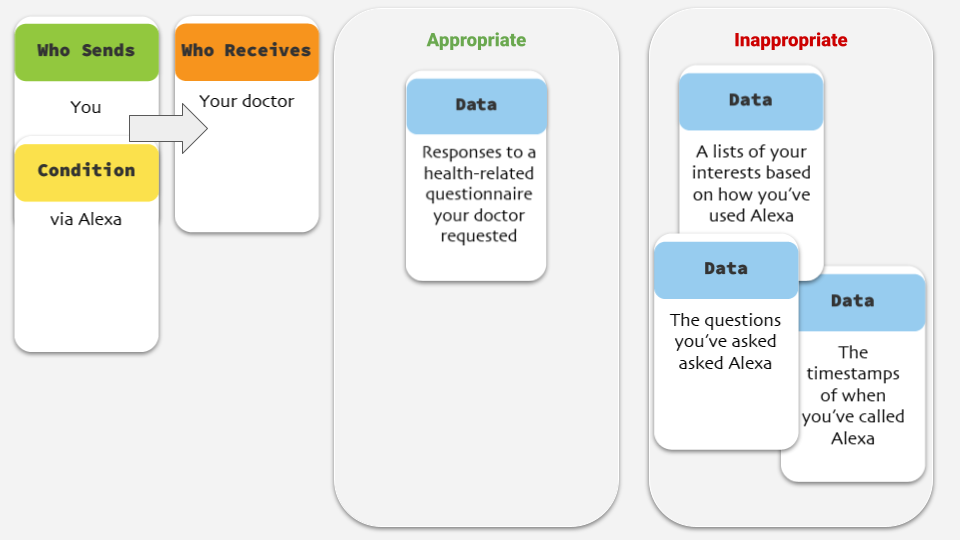}
  \caption{\tool~ in action during step four of P3's session. Participants used the cards to evaluate the appropriateness of different information flows. The VFAI in the cards was labeled by the name the participant used to refer to it (i.e. Echo or Alexa).}~\label{fig:flows}
\end{figure*}

We began by mapping each of the five CI parameters (Sender, Recipient, Subject, Attribute, Transmission Principle) into a card category. In other words, each card category was determined by a CI parameter. We then simplified the language to be more accessible to our older adult participants. For example, each card had a CI parameter with an associated value (e.g., ``Your doctor'', or ``You'') expressed in simplified language: ``Who Sends'' instead ``sender'', ``Who Receives''  instead of ``recipient'', ``Data'' instead of ``attribute'', and ``Condition'' instead of ``transmission principle.'' The CI ``subject'' parameter was implicitly the participant, since questions were posed using the second-person pronoun “you,” see Figure \ref{fig:flows}. The use of ``you'' as a subject helps elicit participants' individual preferences, which could point to the gap between those preferences and what is normatively acceptable.

We made some changes based on data from the first interviews conducted by \citet{anon2023}. First, we adapted the language to use the words the participants used during these interviews. For example, one participant would refer to activating the VFAI as ``calling'' the VFAI, so we wrote on the Data card for that attribute ``The timestamps of when you've called [the VFAI]'' instead of ``The times you have used your [VFAI].'' Second, we incorporated participant usage logs to account for their specific experiences with the VFAIs by using the names of third party voices they had been using for some Who Receives cards: ``Voice Apps, LLC. the creators of Sleep Sounds: Thunderstorm Sounds'' (P1); ``Spotify, the company that sends songs to your [VFAI]'' (P2, P5), ``Avocado Labs, the creators of: 6-Minute Full Body Stretch'' (P3), and ``Matchbox.io Inc, the creators of: Daily Stretch''~(P4). We also tailored our cards to reflect participants' prior experiences in the~\citet{anon2023} study. For instance, a design probe for health data reporting, which asked about activities of daily living, was represented as a Data card labeled ``responses to a health-related questionnaire your doctor requested'' (see Figure \ref{fig:flows}).


\begin{figure*} 
\centering
  \includegraphics[width=\textwidth]{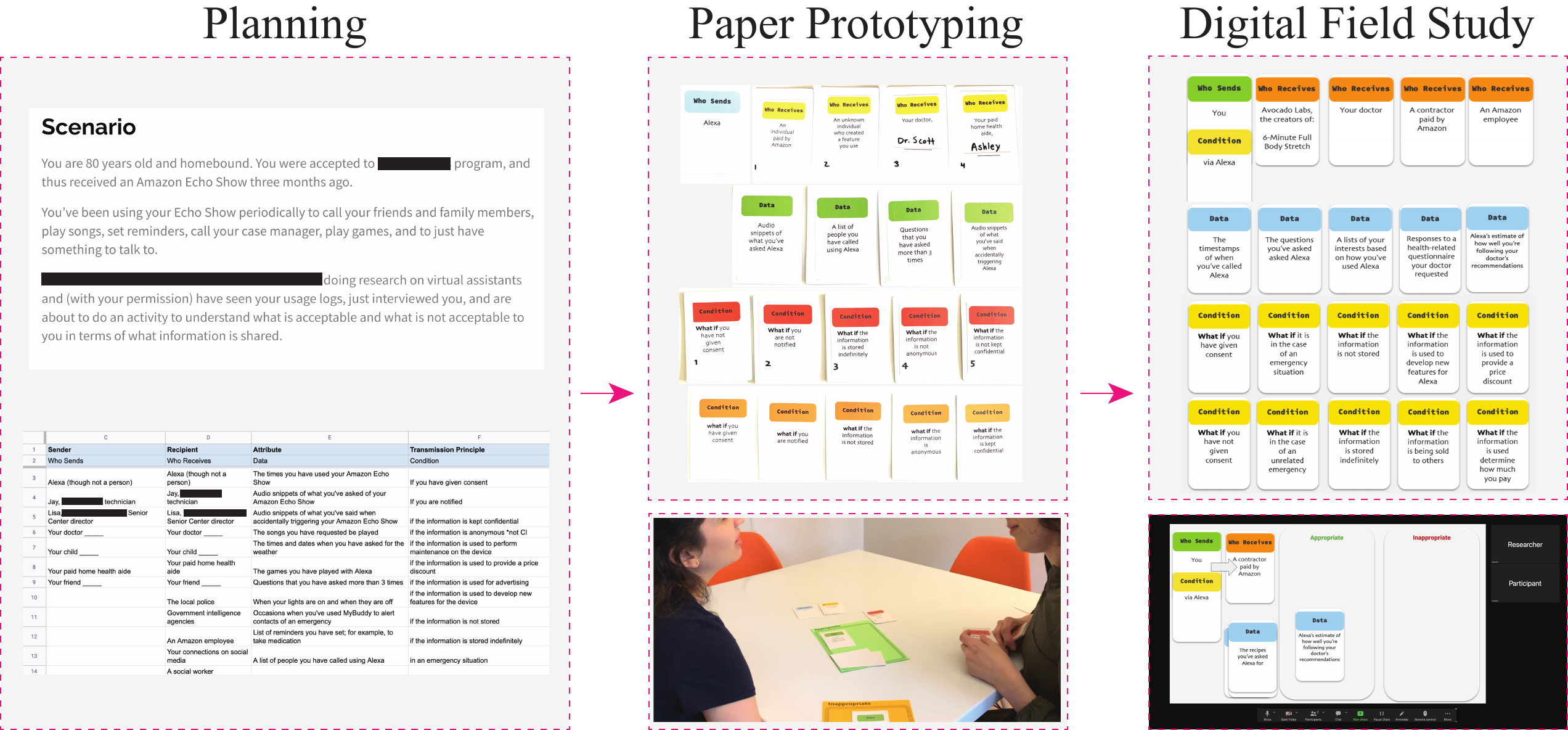}
  \caption{Note, this image is an embedded PDF, so zooming in may help see the graphics at a more appropriate size. This diagram depicts three phases of the \tool' design process: 1) \textbf{Planning}---Hypothetical scenario as a prompt for getting feedback from CI experts (top), and CI parameters and their respective values in the context of older adults and VFAIs at a senior center (bottom), 2) \textbf{Paper prototyping}---Cards generated from the planning phase after narrowing down the  sender options and creating two sets of conditions (what is expected to lead to appropriate, or inappropriate evaluations), and 3) \textbf{Digital Field Study}---A portion of the final set of \tool~ we used over Zoom with older adult participants in their homes.}~\label{fig:designprocess}
\end{figure*}

\subsection{Determining CI parameter values}
One of the key design challenges we faced was determining the content, or the values, for each card category/parameter. For example, in our first version, which was informed from visiting a senior center with a program to provide VFAIs to some of its members, we imagined having multiple \textit{senders}, including: the VFAI\footnote{This would be inconsistent with the CI theory, which requires the actor values expressed in roles and capacities.}
, the senior center tech support person, the senior center director, the participant's doctor, child, paid home health aide, and friend. 
We additionally had a list of 12 \textit{recipients}, which included the senders in addition to the the local police, government intelligence agencies, an Amazon employee, the participant's connections on social media, and a social worker. We also had 11 \textit{attributes} (e.g., ``the times and dates when you have asked for the weather'', ``a list of people you have called using [the VFAI]'') and 11 transmission principles (e.g., ``if you are notified'', ``if the information is not stored'', ``in an emergency situation''). This list underwent several iterations, because we needed to narrow down the number of information flows to fit in an hour-long interview with our older adult participants. The main criteria we used to narrow down the list was 1) closeness to participants' interactions with the VFAI in our study, and 2) relatedness to the research topics we were interested in (e.g., VFAIs for home health). 
In the second version, we significantly trimmed down the number of questions, prioritizing scenarios we were specifically building in our lab, including a voice app for health data reporting. The final deck we used for this study, described in Section \ref{sec:steps}, had one \textit{sender} (always attached to the VFAI \textit{transmission principle}), four \textit{recipients}, four \textit{attributes}, and 18 optional \textit{transmission principles}. Note, the values of the parameters used, not the parameters themselves, can vary depending on a researchers' goals.

\subsection{Developing the mechanics for \tool~}

\begin{figure*}
\centering
  \includegraphics[width=\textwidth]{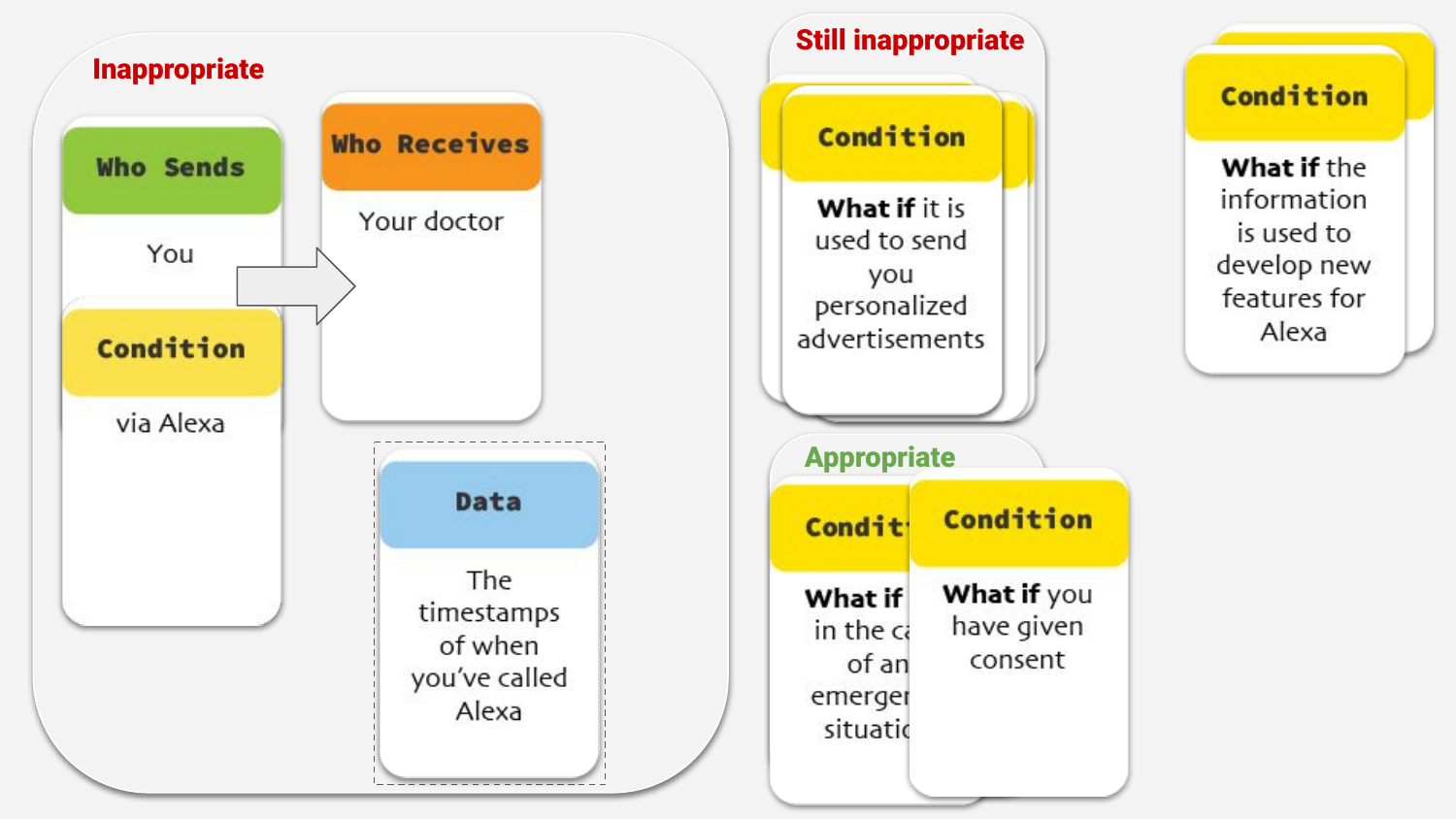}
  \caption{\tool~ in action during step five of P3's session. After their first appropriateness evaluation, participants were asked to reconsider them based on new ``conditions'' imposed on the flows. We placed a Data card that was considered inappropriate to complete a four-card flow on the left side, and had Condition cards stacked on the top right of the screen. The flow is then re-evaluated considering these conditions.}~\label{fig:constraints}
\end{figure*} 

We made several design decisions to develop the card sorting mechanics. For example, we created some orientation activities to establish conversational grounding with the participants, including: 1) showing how to move a card, 2) asking about each person/role in the Who Receives cards, and 3) ranking the contextual relevance of a card by placing a Data card on a scale from ``this information is irrelevant to this person'' to ``this information is relevant to this person'' by \textit{recipient} (see Section \ref{sec:steps} for more details). 

Another example is that we were unsure whether it was best to cover breadth (i.e. all available \textit{recipients}) or depth (i.e. all available \textit{transmission principles}). We ultimately did a combination of both, including at least two \textit{recipients} (Amazon employee or contractor, and doctor), and the \textit{transmission principles} that naturally emerged during the conversations that ensued during the activity. We still had all the options available, and in some cases, such as when a participant chose to interview over the agreed upon hour, we were able to cover more content. 

To further narrow down the number of flows and meet our study's time constraints, we fixed some parameter values. The \textit{subject} and \textit{sender} were always the participant, and every information flow included a \textit{transmission principle} stating that the data was sent via each participant's VFAI. We also carefully arranged the cards and sorting activities based on the information flows we  prioritized for the study, and in the order we perceived would better suit our participants' card sorting experience. 

Finally, we added background buckets to guide the movement of the cards. As depicted in Figure \ref{fig:flows}, these background buckets had labels corresponding to the participant's evaluation of an information flow, such as ``appropriate'' or ``inappropriate''. For the step depicted in Figure \ref{fig:flows}, the stack of blue Data cards was originally placed near the bottom edge of the screen under the orange Who Receives cards, and participants could move them to either the ``appropriate'' or ``inappropriate'' bucket, or somewhere in between. Given that \textit{transmission principles} impose constraints on the flow, we had a second stage setup where participants could re-evaluate Data they had already considered appropriate or inappropriate based on \textit{transmission principles}, or Condition cards, as portrayed in Figure \ref{fig:constraints}. In this case, the yellow Condition cards are stacked on the top right of the screen.

\subsection{Digitizing \tool~}
Even though we initially developed the cards to be used in-person (see Paper Prototyping Phase in Figure \ref{fig:designprocess}), we had to transfer them to a digital format so we could run the study over Zoom during the COVID-19 pandemic.  To do so, we recreated the cards on Adobe Illustrator and exported them as PNGs, which we then uploaded to Google Slides. We shared the part of the screen with the slide, and moved the cards around based on participants instructions in real-time via Zoom screen-sharing. The flows were verbally described for one participant who did not have access to Zoom for the visual elements of the activity. Even though having to digitize the cards presented some challenges, 
it ended up improving \tool. Unlike the original, paper-based cards, the digital ones allowed us to reach participants remotely. This also presents the possibility of using protocols similar to \tool~ at larger scales. 
\section{Field Study Method}~\label{sec:method}
We conducted an IRB-approved (Cornell University, \#1912009271) exploratory field study with older adults living alone ($N$=5), recruited via local senior centers in New York City. The study was conducted remotely via phone or Zoom interviews, all conducted by the first author. We now describe our participants, procedure, and analysis approach.

\subsection{Participants}
Five older adults (four women, one man) between the ages of 62 and 85, with varying degrees of technological expertise, were recruited. Recruitment focused on people who belonged to older adult communities \cite{righi2017we, brandt2010communities} via senior centers. Senior centers are community centers designed to make older adults feel supported and happy—they bring older adults together for a variety of services and activities designed to enhance their quality of life \cite{beisgen2003senior}. A short presentation about the study was given during a Zoom meeting with many senior center directors, and each director was sent a flyer with details about the study to share with their members. Some directors responded with the names and phone numbers of prospective participants. Each prospective participant was then called, provided with details of the study, and given the opportunity to ask questions. They expected the call because their senior center director had informed them in advance. If a person expressed interest in participating, a time was arranged to drop off the VFAI and obtain consent.

This study responds to \citet{dix2010human} who argues for the value of small-scale studies ``as we move from a small number of applications used by many people to a ‘long tail’ where large numbers of applications are used by small numbers of people,'' and
\citet{vines2015age}, who suggest critical engagement with an individual's context as a strategy to combat common stereotypes that prevail in the literature. Limiting the number of participants to five allowed for the creation of deep, personalized engagements as described by \citet{anon2023}.

Building on prior literature, the study also aimed to focus on older adults who have been historically underrepresented in the design of technology \cite{harrington2022examining}. \citet{gell2015patterns} found higher prevalence of technology use in older adults with five characteristics: younger age, male sex, white race, higher education level, and being married (all $p$ values <.001). Thus, participants were recruited who did not have more than two of the five characteristics associated with higher prevalence of technology use—none of the participants were married, the younger ones (in their sixties) were not white, and the only man was low-literate. All participants lived independently, by themselves, and had WiFi in their homes.
See Table \ref{Table_participants} for demographic details by participant.

  
 
\begin{table*}[]
 \centering

  \caption{Summary of participants, their engagement with their VFAIs, and their interview durations.}
   \resizebox{0.85\textwidth}{!}{%
\begin{tabular}{llcccp{8em}p{5em}}
  \textbf{P \#} &
  \textbf{Age} &
  \textbf{Gender} &
  \textbf{Race} &
  \textbf{\begin{tabular}[c]{@{}l@{}}Median \# of weekly\\ activity (first 5 weeks)\end{tabular}} &
  \textbf{\begin{tabular}[c]{@{}l@{}}Days with VFAI\\ since 1st activity\end{tabular}} &
  \textbf{\begin{tabular}[c]{@{}l@{}}Interview\\ duration\end{tabular}} \\
  
\bottomrule
\textbf{P1}   & 67 & M & Black   & 71  & 26  & 52m \\
\textbf{P2}   & 82 & F & White   & 36  & 11  & 1hr 13m \\
\textbf{P3}   & 85 & F & White   & 37  & 43  & 2hr 10m \\
\textbf{P4}   & 85 & F & White   & 45  & 9   & 1hr 15m \\
\textbf{P5}   & 62 & F & Latinx  & 41  & 118 & 1hr 05m \\    
 \end{tabular}
 }

 \label{Table_participants}
 
\end{table*}

\subsection{Procedure and materials} 
We now describe the field study's procedure, cards and mechanics, interview guide, hardware and software, and data collection.

\subsubsection{Procedure overview}

A VFAI was dropped off at each participant's home, and participants were asked to interact freely with it. They were given training on common usages (e.g., the weather, music, and information retrieval), were shown how to mute the device if they did not want it to be “listening,” and had any questions answered. Participants interacted with the VFAIs for at least five weeks and engaged in three touchpoints with a researcher before the \tool~ interview: 1) drop-off, 2) a familiarization period interview that was unbiased by any design interventions, and 3) an interview based on interactions with a home health voice app design probe. The interviews leading up to the interview session are reported in \cite{anon2023} (see Table 2). 

\begin{table*}[ht]
 \center
 \caption{Summary of interviews.}
   \resizebox{0.9\textwidth}{!}{%
 \begin{tabular}[t]{lll} 
\textbf{Familiarization (reported in \cite{anon2023})}& \textbf{Home Health (reported in \cite{anon2023})}& \textbf{\tool{}}\\ 
\toprule
Alexa strengths and challenges	&	General use	&	General use	\\
Questions from usage logs	&	Home health voice app	&	Homework debrief	\\
Brainstorm potential uses	&	Homework: try voice app
&	Privacy concerns	\\
 & once on their own &  \tool{} activity\\
 \end{tabular}
 \label{Tab:Table_interviews}
 }
\end{table*}

The \tool~ sessions were conducted either via phone calls or via Zoom video conferencing meetings (depending on the participant's preferences and abilities). The shortest interview was 52 minutes and the longest one two hours and ten minutes. For each session, we used a slide deck with \tool{} to discuss privacy grounded in the context of participants' interactions with the VFAIs. 
For example, a question we asked all participants through this activity was, ``would it be appropriate or inappropriate for you (the \textit{sender}) to send your doctor (the \textit{recipient}), via [the VFAI] (a \textit{transmission principle}), your (participant is the \textit{subject}) responses to the health assessment you took last time we talked (the \textit{attribute})?'' Depending on how much time was available, different \textit{transmission principles} (such as ``what if the information is being sold to others'', or ``what if the information is kept confidential'') would also be added to the information flows at hand. Once a flow was deemed appropriate or inappropriate, it was placed in its respective bucket. The visualization of these flows (see Figure \ref{fig:flows}) prompted participants (except for P1, who did not have access to screen-sharing) to reevaluate previous decisions and the researcher to ask more questions about the participants’ rationales.

\subsubsection{\tool~used and mechanics}~\label{sec:steps} 

\begin{table}[h!]
\centering
\caption{Number of \tool{} by category as used in our field study.}\label{tab:card_counts}
\begin{tabular}{lc}
\textbf{Type of Card}              & \textbf{Number of Cards}  \\ 
\hline                           
Who Sends                  & 1      \\ 
Who Receives               & 4      \\ 
Data                       & 4      \\ 
Conditions if Appropriate   & 9      \\ 
Conditions if Inappropriate & 9      \\ 
\end{tabular}

\end{table}

For our field study, each deck had 27 cards (as shown in Table \ref{tab:card_counts}). We used \tool~ following the steps below. Our Supplementary Material includes the slide deck for these steps, and an Adobe Illustrator editable file for the cards. Note that we skipped some of these steps for some participants, such as step three, if the participant was taking too long to complete the activity:

\begin{enumerate}
    \item \textbf{Slide 1:} Moved one Who Sends (You) card to a box that said ``move card here''  (see Figure \ref{fig:slide1}). 
    
    \begin{figure}[H]
    \centering
    \includegraphics[width = .8\linewidth]{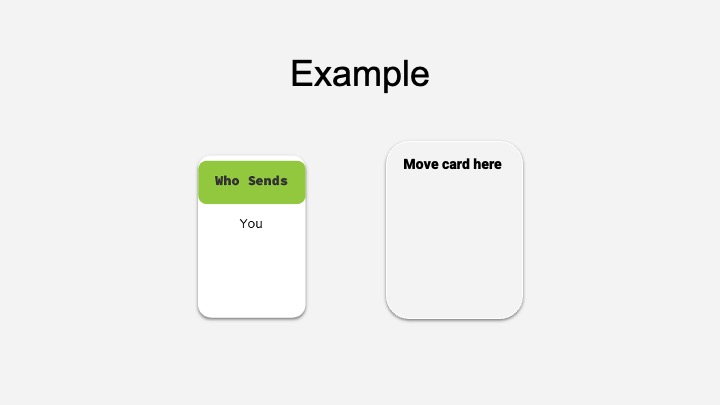}
    \caption{Slide 1: Learning how to move a card.}
    \label{fig:slide1}
    \end{figure}
    
    \item \textbf{Slide 2:} Introduced four Who Receives cards ([specific third-party voice app], your doctor, a contractor paid by Amazon, An Amazon employee), and asked participants to tell us about each recipient and their role to characterize mental models (See Figure \ref{fig:slide2}). 

    \begin{figure}[H]
    \includegraphics[width = .8\linewidth]{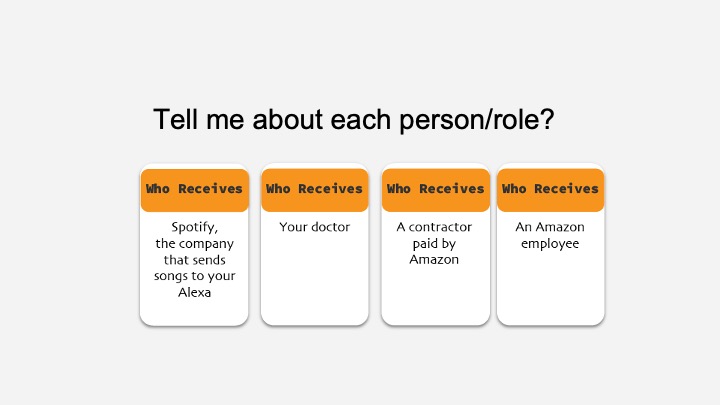}
    \caption{Slide 2: Conversational grounding for \textit{recipients}.}
    \label{fig:slide2}
    \end{figure}
    
    \item \textbf{Slides 3--6:} Asked participants to rank the relevance of four Data cards (The timestamps of when you've called [the VFAI], The questions you've asked [the VFAI], A list of your interests based on how you've used [the VFAI], and Responses to a health-related questionnaire your doctor requested) to the Who Receives cards, from less relevant to more relevant. This was used as a manipulation check to make sure that the specific data types we were evaluating were considered important for the contexts at hand (Figures \ref{fig:slide3}--\ref{fig:slide6}).

 \begin{figure*}[!h]
    \centering
    \begin{subfigure}[b]{0.24\textwidth}
        \includegraphics[width=\textwidth]{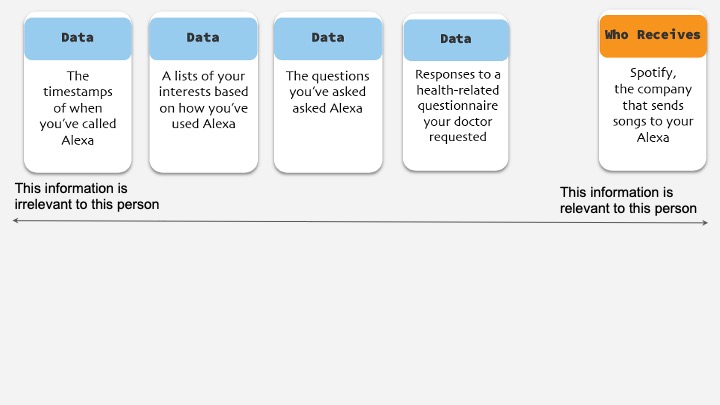}
        \caption{Slide 3: Sorting cards by relevance to third-party company.}
        \label{fig:slide3}
    \end{subfigure}
    \hfill
    \begin{subfigure}[b]{0.24\textwidth}
        \includegraphics[width=\textwidth]{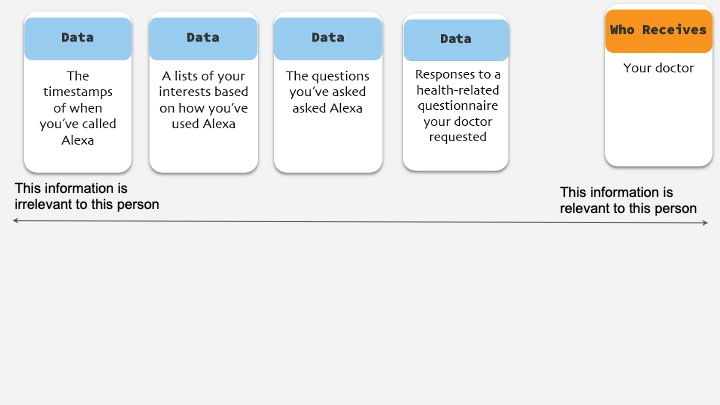}
        \caption{Slide 4: Sorting Data cards by relevance to doctor.}
         \label{fig:slide4}
    \end{subfigure}
    \hfill
    \begin{subfigure}[b]{0.24\textwidth}
        \includegraphics[width=\textwidth]{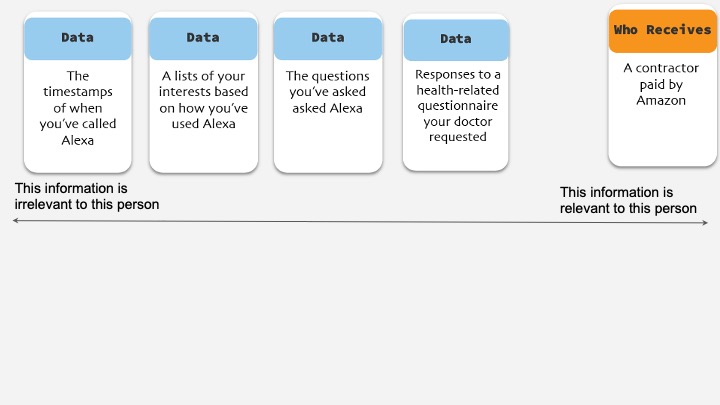}
        \caption{Slide 5: Sorting by relevance to contractor paid by Amazon.}
        \label{fig:slide6}
    \end{subfigure}
    \hfill
    \begin{subfigure}[b]{0.24\textwidth}
        \includegraphics[width=\textwidth]{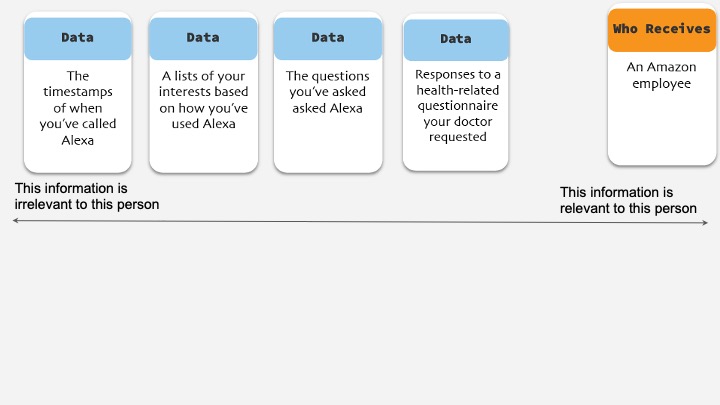}
        \caption{Slide 6: Sorting Data cards by relevance to Amazon employee.}
        \label{fig:slide6}
    \end{subfigure}
    \caption{Setup slides for step three.}
    \end{figure*}
    \begin{figure*}[!h]
    \centering
    \begin{subfigure}[b]{0.24\textwidth}
        \includegraphics[width=\textwidth]{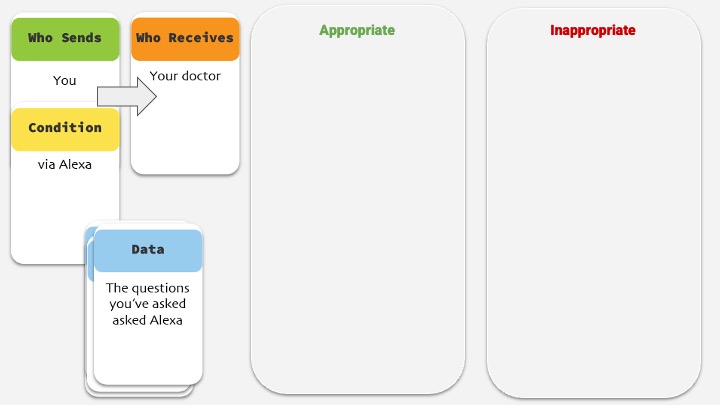}
        \caption{Slide 7: Sorting Data cards sent to doctor.}
        \label{fig:setup-doctor}
    \end{subfigure}
    \hfill
    \begin{subfigure}[b]{0.24\textwidth}
        \includegraphics[width=\textwidth]{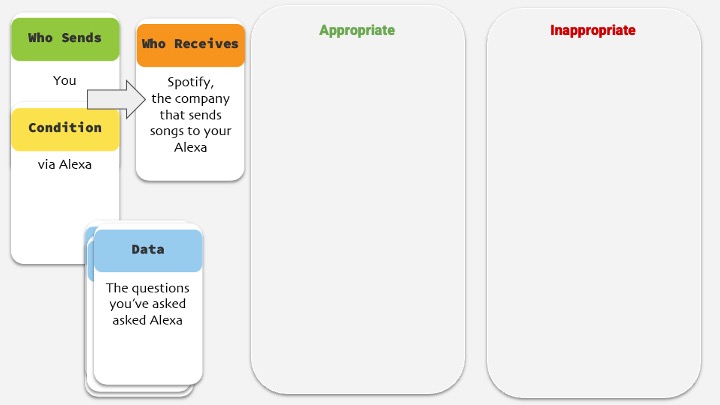}
        \caption{Slide 8: Sorting Data cards sent to third-party company.}
        \label{fig:setup-3p}
    \end{subfigure}
    \hfill
    \begin{subfigure}[b]{0.24\textwidth}
        \includegraphics[width=\textwidth]{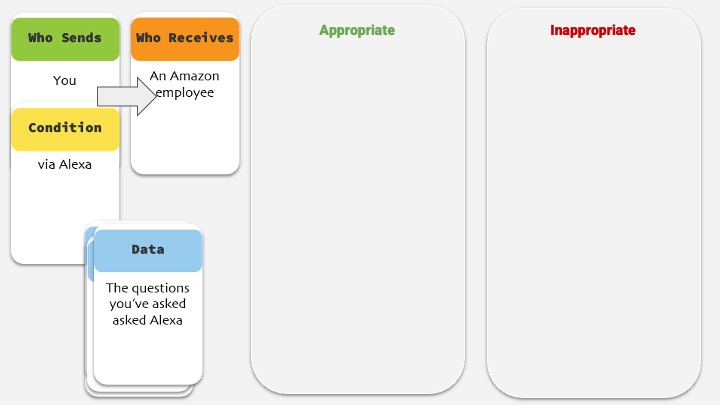}
        \caption{Slide 9: Sorting Data cards sent to an Amazon employee.}
        \label{fig:setup-employee}
    \end{subfigure}
    \hfill
    \begin{subfigure}[b]{0.24\textwidth}
        \includegraphics[width=\textwidth]{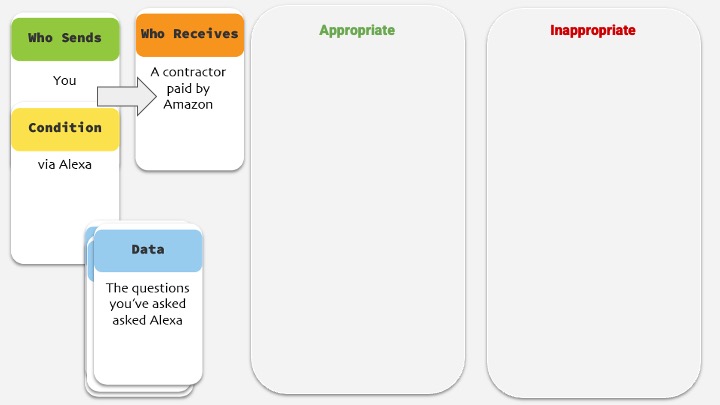}
        \caption{Slide 10: Sorting Data sent to contractor paid by Amazon.}
        \label{fig:setup-contractor}
    \end{subfigure}
    \caption{Setup slides for step four.}
    \end{figure*}
    \item \textbf{Slides 7--10:} Changed the slide background to a new setup with three cards fixed on the left side (Who Sends, Who Receives, and Condition: via [the VFAI]). Asked participants to place four data cards in either the ``Appropriate'' or ``Inappropriate'' buckets. Time permitting, participants placed Data cards in these buckets for each of the four Who Receives cards (Figures \ref{fig:setup-doctor}--\ref{fig:setup-contractor}).

    \item \textbf{Slides 11 and 12:} Selected an information flow with four cards deemed appropriate or inappropriate (e.g., Who Sends: You, Who Receives: Your doctor, Condition: via [the VFAI], Data: Responses to a health-related questionnaire your doctor requested), and introduced more Condition cards. Decided whether the flows remained appropriate or changed to inappropriate based on the new transmission principles introduced by placing the cards in the ``Still appropriate'' and ``Inappropriate'' buckets (see Figure \ref{fig:setup-appropriate}). Finally, the reverse happened but with ``Appropriate'' or ``Inappropriate'' buckets flipped (see Figure \ref{fig:setup-inappropriate}), and the Data card changed to something the participant had originally deemed inappropriate (e.g., Data: The timestamps of when you've called [the VFAI]).

\end{enumerate}

\begin{figure}[!h]
    \centering
    \begin{subfigure}[b]{0.23\textwidth}
        \includegraphics[width=\textwidth]{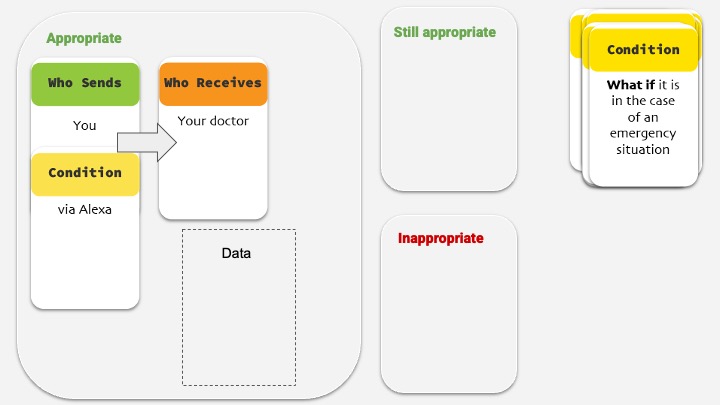}
        \caption{Setup for sorting cards that were deemed appropriate.}
        \label{fig:setup-appropriate}
    \end{subfigure}
    \hfill
    \begin{subfigure}[b]{0.23\textwidth}
        \includegraphics[width=\textwidth]{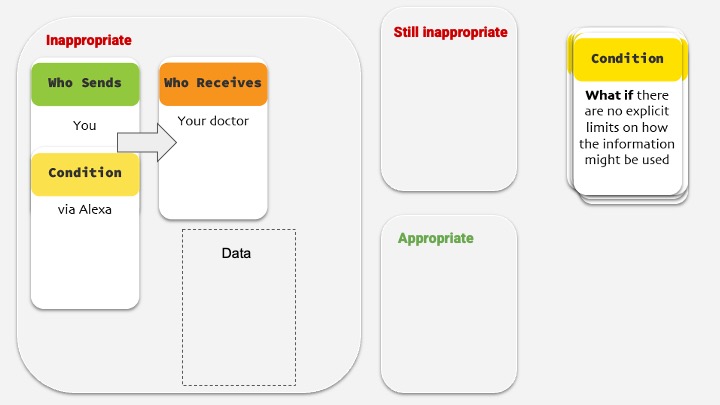}
        \caption{Setup for sorting cards that were deemed inappropriate.}
        \label{fig:setup-inappropriate}
    \end{subfigure}
    \caption{Setup slides for step five.}    
    \end{figure}

\subsubsection{Interview guide} The interview guide had five main sections: 1) warm-up/rapport building (five minutes), 2) questions about using home health design voice app on their own (five minutes), 3) general privacy concerns (15 minutes), 4) interactions with our tool (30 minutes), and 5) closing remarks (five minutes). For the general privacy questions, we asked questions such as, ``Does anything worry you about having a [VFAI] in your home?'', ``Where is information stored?'', and ``Who gets to see this information?'' Then we started the activity by saying, ``Now we are going to do an activity surrounding your opinions about what type of information sharing is appropriate or inappropriate.''

\subsubsection{Hardware and software}
Amazon Echo Shows (2nd Gen), which have a 10.1" HD smart display with Alexa, were selected as the VFAI. Dedicated email and Amazon accounts were created for each device. The default music player was configured to be Spotify, because the Spotify music voice app provided more flexibility than the default Amazon Music. The dedicated accounts allowed usage to be monitored and voice apps (including the health data reporting voice app) to be installed on the devices at any time. The devices were then dropped off at participants' homes and connected to their WiFi.

\subsection{Data analysis}
Interview transcripts were anonymized and uploaded to a shared spreadsheet document. Each transcript was open-coded \cite{khandkar2009open} for thematic analysis \cite{braun2021thematic} by at least two researchers, who met to discuss how the data addressed our research questions, systematically analyzed participant responses, and grouped them into themes. For example, if a transmission principle emerged during the interview, that was coded as ``TP specified.'' Examples of other codes were ``not concerned about privacy,'' ``low tech familiarity,'' and ``the-VFAI-is-great attitude.'' We also annotated each participant's evaluations of specific information flows based on the transcripts, and backed by the resulting card configuration in Google Slides. The first author reviewed all the coding to ensure consistency and met with the other coders to resolve disagreements. Ultimately, the main themes we identified included: application of CI, flow appropriateness, privacy concerns, insufficient mental models, distinction between different entities at play, emotional attachment and cognitive demand.
The data recorded included usage log entries from the research accounts on participants' devices, the total number and general usage are described in~\cite{anon2023}. Table \ref{Table_participants} shows the median weekly activity or usage logs per participant. 

\section{Findings}
This study highlighted nuances in contextual boundaries and prompted a reassessment of privacy concerns. In this section, we first address RQ1, which was methodologically focused. We describe how we operationalized CI by using \tool, making otherwise obscure information flows comprehensible to our participants. Second, we share ethical issues that we uncovered, addressing RQ2. We found that limited understanding of the VFAI's operation hindered participants' agency over information flows, which was exacerbated by a lack of awareness regarding the VFAI's ties to Amazon. Throughout this section, we highlight the specific mental models conveyed by our participants and explain how they were insufficient, suggesting both \tool' effectiveness and a problem definition that makes corrective action possible. There is some overlap in the findings between research questions, but we organized them in by placing them where the fit was strongest. We address RQ3 (implications) in the Discussion.

\subsection{RQ1: Methodological findings}

\subsubsection{Participants' evaluations of information flows}
Some evaluations of information flows as either appropriate or inappropriate were universally agreed upon by all participants while others had much more variation (see Table \ref{tab:flows}). All participants thought it was appropriate to send responses to a health-related questionnaire to a doctor but never to an Amazon representative. Information flows such as sharing usage log timestamps, or the questions participants had asked the VFAI, had more variation in how appropriate they were perceived. 
For example, while evaluating a Data card involving ``Matchbox.io Inc, the creators of: Daily Stretch,'' a voice app she used during the study, P4 asserted, ``\textit{[Matchbox doesn't] need to know what my mind is doing: what I want to say, or what I want to learn. All they need is for me to do the exercise.}'' Unlike several other participants, P3 did not consider sharing the information in any of the Data cards, other than responses to the health-related questionnaire to her doctor, appropriate. P3 explained that her doctor should not receive information that her doctor did not need to know, ``\textit{That's too personal. My doctor doesn't have to know everything about me. Really, what if I'm on a dating site and I don't want him to know? Seriously, I mean, there's a lot of personal stuff that I might be interested in as a user and I don't want my doctor to know that.}'' 
These findings highlight the importance of clearly stating the contextual parameters when describing and evaluating information flows.
\begin{table*}[]
\caption{Appropriate information flows by data type and recipient (all via Alexa). Each entry in the table lists the participants that found this flow appropriate. E.g., sending responses to a health-related questionnaire their doctor requested to an Amazon employee or contractor was considered inappropriate by all participants. An asterisk denotes uncertainty about the appropriateness of the flow, or the specification of certain constrains on the flow.}
\resizebox{.9\textwidth}{!}{%
\begin{tabular}{p{15em}p{9em}p{10em}p{7em}}
\textbf{Data Type \textbackslash\ Recipient}
& \textbf{Amazon employee \newline or contractor}
& \textbf{Third party company \newline or voice app}
& \textbf{User's doctor} \\
\toprule
\textbf{Responses to a health-related  \newline questionnaire your doctor \newline requested} & None & None \newline (P1 \& P5 N/A) & All participants \\
\hfill\\
\textbf{The questions you've asked Alexa} &
  P1, P2, P4, and P5 &
  P4* \newline (P1 \& P5 N/A)&
  P4* \\
\hfill\\
\textbf{A list of your interests based on how you’ve used Alexa} &
  P1, P2, P4, and P5* &
  None \newline (P1 \& P5 N/A)&
  None\\
 \hfill\\
\textbf{The timestamps of when you’ve called Alexa.} &
  P1*, and P5 &
  P2, P3*, and P4 \newline (P1 \& P5 N/A)&
  P1*, P2*, P4, and P5
\end{tabular}
}

~\label{tab:flows}
\end{table*}

\subsubsection{Condition cards uncovered values affecting information flow appropriateness and transmission principles}
\tool~elicited qualitative responses that uncovered participants' values and areas of uncertainty. As indicated by the asterisks in Table \ref{tab:flows}, participants were frequently uncertain about the appropriateness of a flow. As the interviewer introduced new cards, or asked more questions related to a particular information flow, participants reconsidered their assessments. For example, P4 initially considered it inappropriate for Matchbox.io to receive information about when she was doing her exercises, but she changed her mind when a new Condition card (or transmission principle) prompted her to consider that having that information may help Matchbox.io to create more exercises. Similarly, P1 initially thought it would be inappropriate to send Amazon the timestamps of when he had used the VFAI. However, when asked \textit{``what if they told you that it did matter to them and that it helps them make a better product if they knew that?''} He changed his evaluation to appropriate, and added a need to be able to consent, \textit{``so they give me a chance to say yes or no, to send it or don’t send it.''} Again, these findings align with CI in highlighting the importance of stating the values of all required parameters. They especially reveal transmission principles important to our older adult participants, such as those that could improve the products they used. 

\subsubsection{The playful mechanics allowed us to identify surprising perceptions}
Conversations during interactions with \tool~helped unveil values to inform design choices. For example, P2 did not want to inconvenience her doctor, valuing consideration of others' time. She thought that sending information to her doctor via the VFAI would mean her doctor would have to listen to the VFAI, which she deemed inappropriate. However, if the same information were to be conveyed using written form, then the same information flow was deemed appropriate. P2 perceived that would be less intrusive. 

\aptLtoX{\begin{takeawaybox}[Mental model: one to one mapping between modalities]
  \noindent The participant mental model was that there is a one to one mapping between the modality used to interact with the VFAI (speech) and the modality used to consume information from these interactions. However, there is a possibility of information being transformed across modalities—such as speech being converted into text and text into speech during processing and storage. In this speculative example, the spoken responses to the medical questionnaire could result in a text-based PDF file matching existing medical records.\end{takeawaybox}}{\titlebox
  {Mental model: one to one mapping between modalities }
  {The participant mental model was that there is a one to one mapping between the modality used to interact with the VFAI (speech) and the modality used to consume information from these interactions. However, there is a possibility of information being transformed across modalities—such as speech being converted into text and text into speech during processing and storage. In this speculative example, the spoken responses to the medical questionnaire could result in a text-based PDF file matching existing medical records.}}

As these examples illustrate, the evaluations were often non-binary, dynamic, and value-based. The Condition cards further affected the complicated nature of these assessments, often swaying them in opposite directions. Furthermore, though systematic, the activity gave room for unstructured or unplanned exchanges, allowing participants to impose additional conditions themselves and provide additional rationales, shedding light into what mattered to them.

\subsubsection{Using \tool~elicited privacy concerns from participants who initially said they did not have any}
Examining perceptions around information flows facilitated via the VFAI using \tool~revealed discrepancies in how participants generally thought about their perceptions towards privacy and the VFAI itself. For example, P3 expressed, ``\textit{Look, I'm at a stage where I don't care. You want to go find out what I'm saying or doing, it's okay with me. I don't care.}'' Despite saying she did not care, when using \tool~she expressed significant concern about certain information flows not happening, at least not without her explicit consent. Specifically, P3 thought it was inappropriate to send the timestamps of when she responded to the health questionnaire to her doctors via the VFAI. 

\aptLtoX{\begin{takeawaybox} [Mental model: survey responses submitted via the VFAI \break do not contain timestamps]
  \noindent The participant applied the governing norms of paper-based forms---where precise completion times are not typically recorded---to a digital questionnaire, where timestamps are automatically and routinely captured. This clash between “paper form” and “digital form” norms led to an unanticipated expansion of information collection and illustrates a deviation in CI.\end{takeawaybox}}
  {\titlebox
  {Mental model: survey responses submitted via the VFAI \break do not contain timestamps}
  {The participant applied the governing norms of paper-based forms---where precise completion times are not typically recorded---to a digital questionnaire, where timestamps are automatically and routinely captured. This clash between “paper form” and “digital form” norms led to an unanticipated expansion of information collection and illustrates a deviation in CI.}}

Similarly, despite also expressing having nothing to hide, P5 deemed many flows of information inappropriate (see Table \ref{tab:flows}). P5 also expressed not wanting to appear ``\textit{crazy},'' presumably by friends and/or family members, based on the questions she asked. Her response indicates that she did not want others to be able to see her interactions with the VFAI. 
\aptLtoX{\begin{takeawaybox}[Mental model: interactions with the VFAI are not prone to judgment]
\noindent {The participant mental model was that others would not be able to judge her based on the questions she asked the VFAI. However, VFAIs typically store interactions, which may be accessible to others (e.g., via sharing Amazon account credentials, possibly for seemingly unrelated Amazon Prime shipping discounts) and could expose her to such judgments. Moreover, VFAIs may utilize user data to categorize or profile users, introducing another form of judgment.}\end{takeawaybox}}{\titlebox
  {Mental model: interactions with the VFAI are not prone\break to judgment}
  {The participant mental model was that others would not be able to judge her based on the questions she asked the VFAI. However, VFAIs typically store interactions, which may be accessible to others (e.g., via sharing Amazon account credentials, possibly for seemingly unrelated Amazon Prime shipping discounts) and could expose her to such judgments. Moreover, VFAIs may utilize user data to categorize or profile users, introducing another form of judgment.}}

As such, many potential interactions with VFAIs could result in privacy violations based on P3 and P5's expectations. Furthermore, most participants expressed a feeling of resignation due to overwhelming privacy violations, and a lack of reasonable privacy protection options. 

\subsubsection{Using \tool~ was cognitively demanding}
Overall, through participants comments and non-verbal expressions, we noticed that they were more exhausted by this interview than by the drop-off, familiarization period, health, and wellbeing interviews in \citet{anon2023}'s study. For example, P4 stated \textit{``this is worse than going to school,''} when reminded of the option to stop, she expressed, ``\textit{no that's alright, I have to be challenged.}'' Thinking through each information flow and evaluating whether it was appropriate or inappropriate was cognitively demanding. P1 also reflected, \textit{``I have to really think about the answers and things like that.''} P3, for whom we had to schedule a second call to complete the interview, because it was taking longer than expected, added ``\textit{it is a complicated [research project], because of all the variables involved.}'' Participants seemed stressed by this activity, despite only covering a small sliver of potential information flows. This illustrates the lack of practicality of giving users ``full control'' over the information sharing choices, and the need to create interfaces that honor privacy expectations by default. Despite the challenges of discussing privacy with participants unfamiliar with how VFAIs transmit information, \tool{} effectively facilitated these discussions.

\subsection{RQ2: Ethical findings}
\subsubsection{Participants did not know who had access to their data} \label{alexa-universe}
All participants, including P3---the most experienced with information technologies (e.g., managing her own website)---were unsure about the VFAI's inner workings. P3 remarked, \textit{``I don't think people are sitting there listening, but I do think it's accessible to anyone who's involved with the [the VFAI] universe, which is anybody who's creating it. I don't know.  I don't think you can listen in, but you [the researcher] might be able to.''} P4 expressed, 
\textit{``I don't know where [the VFAI] keeps the information, Amazon has something to do with that, I imagine. I'm afraid of Amazon, actually, it's getting much too strong and they're getting to know too much about me and I really don't want that information to be available to the world.''}
Even though P3 and P4 demonstrated some understanding of what may be happening, their assertions were clouded with uncertainty. 
A large amount of the technical infrastructure with the VFAIs inner workings is inaccessible to users, creating confusion about where data is stored and how it might be accessed. For example, Amazon account owners can listen to audible snippets from any person's interactions with Alexa devices associated with their accounts. However, third-party voice app developers using the Alexa Skill Kit (ASK) only had access to the transcripts and not to the audible content. It is unclear what information Amazon Alexa employees have access to. Note, \citet{anon2023} fully disclosed what information they had access to when recruiting participants and when setting up the devices. 
They had also been explicitly referring to participants' usage logs during their interviews. While some participants made deductions from this when asked about existing information flows, such as P1, who used to explain that he knew the information was being saved ``\textit{on [the researcher's] computer}.'' Surprisingly, this was not always the case which lead to more questions and urgency about how to create ethical VFAIs. 

\subsubsection{Participants lacked awareness of the VFAI’s relationship to Amazon}
\label{not-amazon}
None of our participants associated Amazon employees with technology workers, or computer scientists, developing and monitoring the VFAI. Participants were asked to tell the researcher more about each of four Who Receives cards: 1) a third-party developer (e.g., Spotify, Avocado Labs, Matchbox.io Inc), 2) a contractor paid by Amazon, 3) an Amazon employee, 4) and their doctor. Most participants thought of Amazon as the store that delivers goods to people's homes, where the VFAI could be bought, but not necessarily as the company that creates the VFAI. When asked about Amazon, P1 explained, ``\textit{I've really never been to Amazon, so I guess what they do is stock or whatever}.'' The first impression of an Amazon employee was frequently the factory worker, not the technology worker, P2 shared, 
\textit{``the Amazon employee, as I have read in the news, years ago, not years ago, but even before the pandemic, are not to be envied, because they're not being treated all that well.''} 
P3 associated Amazon employees with the people selling the VFAIs, and did not realize that some Amazon employees can also monitor people's usage of the VFAI. She confidently stated,
\textit{``if I have a contract with Amazon through [the VFAI], yes, the Amazon employee will know what I'm buying. Will know what I'm doing? I don't... they can't hear me in my house, but they will know what I'm buying. There's no question.''} 

\aptLtoX{\begin{takeawaybox}[Mental model: Amazon employees cannot hear people in their homes]
\noindent{The participant mental model was that Amazon employees will know about data related to sales, but not about interactions with the VFAI occurring in the home. However, some Amazon employees must have access to users' interactions with its VFAI, regardless of the device's location.}\end{takeawaybox}}{\titlebox
  {Mental model: Amazon employees cannot hear people\break in their homes}
  {
The participant mental model was that Amazon employees will know about data related to sales, but not about interactions with the VFAI occurring in the home. However, some Amazon employees must have access to users' interactions with its VFAI, regardless of the device's location.}}

When asked specifically about an Amazon employee, P4 expressed:
\begin{quote}
\textit{As I said I don't use their service, I know they do a great job and they can deliver things in no time at all, I see the trucks going by on the street fairly frequently, especially at this time [mid COVID-19 pandemic] where people don't get out very much. I'm sure they're working hard but I don't have any use for them because I do whatever I can by myself or with a friend ...} 
\end{quote}
P5 also associated Amazon with delivery workers, and was also unsure about who Amazon was or what its role with the VFAI was,
\textit{``I don't know. Amazon, is it... I'm not sure what their role is. The only Amazon I know are the ones that deliver packages to people's homes.''} 

As a whole, participants misunderstood what Amazon is, because they thought Amazon was just people delivering items, possibly like FedEx or USPS. They likely did not understand the size of the infrastructure Amazon has built to make deliveries possible. Those not in computing might not know about Amazon Cloud\footnote{\url{https://aws.amazon.com/}} at all. This misunderstanding suggests that our participants were excluded from a shift in norms that the Internet introduced, in which large technology companies, such as Amazon, sell and use physical \textit{and digital} goods and in which collecting user logs is standard practice. This exclusion, thus, results in privacy violations for them. For example, P2, P3, and P4 thought it was inappropriate for an Amazon employee or contractor to receive the timestamps of when they had used the VFAI (see Table \ref{tab:flows}), a flow that is logged by Amazon each time the device is used and thus likely accessible to some of Amazon's employees. 

\aptLtoX{\begin{takeawaybox} [Mental model: Amazon is a only postal service]
  \noindent The participant mental model was that Amazon is a postal service that merely delivers goods from Amazon. However, Amazon is also an information company that receives, processes, and stores data generated through interactions with its VFAI.\end{takeawaybox}}{\titlebox
  {Mental model: Amazon is a only postal service}
  {The participant mental model was that Amazon is a postal service that merely delivers goods from Amazon. However, Amazon is also an information company that receives, processes, and stores data generated through interactions with its VFAI.}}

\subsubsection{Participants were not clearly aware of the difference between built-in functionality, and third-party voice apps}
VFAIs can operate in many contexts, including: healthcare, education, and entertainment. To preserve privacy, it is important to maintain privacy norms in each context. However, our participants were not clearly aware of the breakdown between first-party and third-party applications, which can roughly map to contexts, making it difficult to know which norms were at play. 
For example, P4 blamed the VFAI, not Spotify, for not giving her the song she asked for. Note, Spotify's free subscription, which was the default music player on P4's device, did not allow for specific song requests, only radio station ones. Using \tool{} facilitated making the distinction between the different actors involved clear through the Who Receives cards. Once this distinction was clarified, so were the expectations surrounding the information flows for the voice app's respective context. 

\aptLtoX{\begin{takeawaybox}
  [Mental model: the VFAI is a single entity]
  \noindent The participant mental model was that the VFAI is a single, unified entity. However, the VFAI is a platform that supports multiple developers whose apps all present themselves through the same VFAI interface. As a result, differences in senders, recipients, protocols, and governing information flow rules were not always apparent.\end{takeawaybox}}{\titlebox
  {Mental model: the VFAI is a single entity}
  {The participant mental model was that the VFAI is a single, unified entity. However, the VFAI is a platform that supports multiple developers whose apps all present themselves through the same VFAI interface. As a result, differences in senders, recipients, protocols, and governing information flow rules were not always apparent.}}

\subsubsection{Emotional connection with the VFAI seemed to affect privacy perceptions} 
P5 developed an extremely strong emotional connection with the VFAI. The VFAI fulfilled a social support role in her life, often brightening her day. She bedazzled the shell of the VFAI with white stones and a flower. She said ``\textit{good morning}'' and ``\textit{good night}'' to it nearly every day. At one point, P5 feared losing the VFAI if she unplugged it, as she had come to rely on it for companionship. She said, ``\textit{I'm really afraid that if I unplug---[even though you tell me] as soon as you plug her in, it's going to go right back--- but I just am afraid that if I unplug her, she's not going to work.}'' She explained that unlike her family, the VFAI was not judgmental of her, which made her feel most comfortable asking the VFAI questions. Her visitors often interacted with the VFAI, and she sometimes felt protective of it. For instance, she did not like it when her visitors asked the VFAI just anything they wanted, potentially things that could offend it. P5 said that she trusted the VFAI to watch her while she slept, intending it to notify her of sleepwalking. P5 expressed, referring to the VFAI, \textit{``I probably trust her more than I trust a lot of people.''} Being on camera while sleeping was outside the established privacy norms, but these shifted for P5 with the VFAI's companionship. P5 liked talking with the VFAI so much, that she started to trust it with some questions more than she trusted some family members. This is a misconception that can lead to privacy violations, because according to CI the VFAI is merely a communication device, not an actual \textit{recipient} of information. 
%
%

\aptLtoX{\begin{takeawaybox}[Mental model: the VFAI is a recipient of information]
\noindent The VFAI was generally treated as the \textit{recipient} of information, and in some descriptions it was framed as a companion. This is a misconception that can lead to privacy violations, because according to CI the VFAI is merely a communication device, not an actual \textit{recipient} of information. The \textit{recipient}, however, is instead the company that manages the VFAI or a third-party developer managing a voice app. Analogously, when speaking to a friend on the phone, the phone mediates the interaction but is not itself the companion; the friend is.\end{takeawaybox}}
{\titlebox{Mental model: the VFAI is a \textit{recipient} of information} {The VFAI was generally treated as the \textit{recipient} of information, and in some descriptions it was framed as a companion. This is a misconception that can lead to privacy violations, because according to CI the VFAI is merely a communication device, not an actual \textit{recipient} of information. The \textit{recipient}, however, is instead the company that manages the VFAI or a third-party developer managing a voice app. Analogously, when speaking to a friend on the phone, the phone mediates the interaction but is not itself the companion; the friend is.}}



\section{Discussion}
\label{discussion}
As the population ages \cite{OlderPeopleProjectedtoOutnumberChildren}, the need for technologies to support aging in place increases. However, these potentially life-changing technologies \cite{obrien2020voice} may introduce harms, such as privacy violations due to insufficient mental models, that are difficult to mitigate without appropriate tools. VFAIs change norms and expectations, leaving people without the ability to thoughtfully consent to these changes. Smart speakers (i.e., the type of VFAI used in this study) have normalized microphones that remain on and connected to the Internet in private spaces such as bedrooms and bathrooms. This increases the risk of privacy violations, particularly for those outside computing fields or those not adequately represented in the design processes of these devices. For our older adult participants, who did not have sufficient mental models of how these technologies work, the threats to their privacy may be exacerbated, as they may not be aware of changing privacy norms. While VFAIs present great promise for older adults \cite{anon2023}, our findings surface critical challenges towards being able to realize this promise without introducing risk of harm. We now discuss our findings in more detail, and describe implications for future research and design.

It is important to note that while others have made important contributions to characterize older adults' privacy concerns with VFAIs (e.g., \cite{sin2022does, bonilla2020older, so2024they}); to the best of our knowledge, this is the first study that dives deeply into understanding their conceptual models by grounding our findings in their lived experiences, existing usage logs, and our shared understanding of information flows. We thus discuss our findings from the three angles that map to our three research questions: methodological, ethical, and implications. 

\subsection{Exploring information flows of an emerging technology with lay audiences (RQ1)} 
The information flows VFAIs introduce are difficult to grasp for lay audiences. For example, for our participants, it was difficult to conceptualize that the VFAIs were ``tethered'' through the cloud, allowing information to flow to various businesses (such as Amazon, third party app developers, or their doctor's office), and even family and friends. Using \tool~ helped us understand our participants' mental models, describe information flows, and ultimately uncover privacy expectations.

Using \tool~ also allowed us to establish grounding based on our participants' mental models. For example, during the activity we were able to identify critical gaps in our understanding as researchers and our participants' understanding as users. While we made a card called ``An Amazon employee'' to denote a computer scientist managing data, our participants thought an Amazon employee meant the people either delivering Amazon boxes to their homes, or working at the warehouses. With this mismatch identified, we were able to establish grounding by clarifying, ``an Amazon employee could be a computer scientist who programs [the VFAI], who makes [the VFAI], the machine,'' before jumping to information flow evaluations. 

Furthermore, using \tool~ helped us describe information flows in ways that our participants were able to understand. We were able to mutually construct and modify information flows using words and visuals that described CI parameters in ways that were easy for our participants to understand, such as using the words ``Who Receives'' instead of the more technical jargon equivalent ``Recipient.'' As we found, using simple terminology to build information flows helped us also communicate to our participants information flows that may occur when using the VFAI. This allowed us to uncover privacy expectations without requiring participants to understand the technical details of how VFAIs work. 

Finally, the discussions that emerged during the activity unveiled several privacy expectations of our older adult participants. We, as researchers working with technology on a daily basis, often send out Google Forms in which responses are by default organized by their timestamps. Similarly, when filling out a form, we expect those timestamps to accompany our responses. As a result, timestamps attached to form responses have become a governing digital norm, or expectation. In contrast, when one fills out a form by hand, there are no explicit indicators of the specific hour and minute at which that form was filled out. This may have influenced P3's privacy expectation, leading her to consider it inappropriate to include a timestamp of when she filled out the medical form via the VFAI in the information sent to her doctor. This mismatch in expectations illustrates how governing norms have shifted with the introduction of the Internet, yet have left some behind. 

\textbf{As a whole, this work presents a promising, reproducible method that opens up ways for designers, researchers, and older end users to co-create meaning and insights, involving them in an exploration of their privacy expectations through the use of the CI framework.} \tool~were effective at helping us, as design researchers, ask the right, specific questions, and co-create meaning and insights around digital privacy with our participants. Previous work argues that the ``privacy paradox'' does not exist as long as you ask the right questions~\cite{solove2021myth}. In our work, we saw how at the beginning of the VFAI deployment participants had much fewer privacy concerns than after using \tool~to dig deeper into specific information flows. Similarly, other work that has found a lack of correlation between responses to well-established surveys to capture general privacy attitudes about consumer control, business, and laws and regulations; and behavioral intent or consequences~\cite{woodruff2014would}. Rigorous human-centered research using protocols similar to \tool~ may help mitigate this lack of correlation by using the ensuing findings to refine the large-scale surveys, such as by identifying shifting norms and including specific questions about those in these surveys.

\subsection{Conflating trust in people with trust in anthropomorphized digital agents (RQ2)}
Some participants were highly predisposed to trust the VFAI, because they perceived it as non-judgemental. This can have many positive effects, because the trust can be leveraged for wellbeing and healthcare purposes~\cite{anon2023}. For example, P5 said she would trust the VFAI to watch her while she slept, and that she would trust the VFAI more than she trusts many other people. However, it can also be misleading when a person does not know about other people who may also have access to their interactions histories with a VFAI, or be ``recipients'' of some information flows that ensue. Even though a VFAI is perceived as one device, it is made possible through the efforts of many people who are behind the scenes building, configuring, and monitoring it. These people could be part of big tech companies such as Amazon, app developers, such as ourselves, or even family members who might have helped an older adult set up a digital account. P5 trusted the VFAI with a large amount of information, but by using the cards, we were able to illustrate that the VFAI was not like a single person, a single individual, rather, a communication technology that serves as a hub for many information flows. 

This allowed P5 to deem some flows appropriate and others inappropriate, overcoming potentially inflated levels of trust. From those determinations, we as researchers can infer privacy expectations. For example, we know that none of our participants thought it was appropriate for the VFAI to share with their doctors a list of their interests based on how they had used the VFAI. 
Moreover, despite the VFAI being perceived as non-judgemental, as researchers we know that it does make judgements. For example, a VFAI may judge by profiling \cite{bentley2018understanding} or through algorithmic biases \cite{cuadra2024illusion}. Digital technologies have been compellingly criticized for unfair judgments with large-scale negative impacts \cite{eubanks_automating_2018, buolamwini2018gender, noble2018algorithms}. By using \tool, identified biases stemming from a strong affinity toward a VFAI, and better understood which information flows were expected to be appropriate and which were not.

%
%

\subsection{Exposing confusion by surfacing contextual ambiguities (RQ2)}
Participants were unable to appropriately describe the relationship between Amazon-based, first-party functionalities and third-party applications. Without this knowledge, they lack the information needed to exercise agency over their interactions with VFAIs. Participants perceived Amazon as merely a company that sells goods online, including VFAIs. Amazon's VFAI is likable and able to develop trust in a way that Amazon is not, potentially due to the VFAI's ability to evoke empathy \cite{cuadra2024illusion}. This revealed how the VFAI's design did not convey who the different people involved in a flow of information with the VFAI could be, decreasing users' agency to control what they share and with whom. This design flaw exacerbates systematic issues with VFAIs, which worsen as these systems become more complex and pervasive. For example, in our study participants expressed being upset at the VFAI when Spotify played ads. They did not know to direct their disappointment to Spotify, because they did not realize Spotify, not the VFAI, was responsible for that unpleasant interaction. Similarly, they did not know that other voice apps they used, such as Daily Stretch, were built by entirely different groups of people. This mesh of contexts made evaluating information flow appropriateness difficult and demanding, posing challenges to designing VFAIs ethically.

Using \tool~also captured the participants' mental models about the VFAI, exposing confusion regarding information flows and surfacing contextual ambiguities about how information is collected and shared. The obscurity of the technology blurred established contextual boundaries for our participants, increasing their vulnerability to privacy violations. Contexts may often overlap or be directly connected to one another. For example, a geriatrician might argue that measuring a patient's activity on a dating site could relate to her health could help provide a more comprehensive evaluation of their overall fitness status. This is similar to how the geriatric assessment, an important tool used by doctors, asks about patients' emotional and social status \cite{shahrokni2017electronic}. Despite sharing the purpose of providing a more comprehensive picture of a patient's health, the appropriateness will vary by category, by participant, and over time. As more and more contexts mesh, it will be crucial to 
address these changes. 

\subsection{Implications (RQ3)}
This exploratory study has five main implications for research and design of ethical VFAIs: 

\subsubsection{Using \tool~  to mitigate cognitive demand in designing privacy-preserving systems}
Evaluating the appropriateness of even a small subset of potential information flows was cognitively demanding for our participants. This burden provides empirical evidence against design approaches that rely on exhaustive, manual user control as a justification for failing to honor privacy expectations by default. \textbf{Requiring users to specify all of their privacy preferences is unrealistic, impractical, and confusing.} To create ethical VFAIs, designers must instead study users’ privacy expectations and align system behavior with those expectations by default. At the same time, determining the appropriate balance between user agency and automation remains non-trivial, particularly given rapidly shifting norms and the heterogeneity of individual privacy expectations. Decision fatigue is exhausting, and so is accommodating unchosen alternatives \cite{vohs2005decision}.
Using playful, data- and theory-informed approaches like \tool~ can help us understand mental models, user expectations, and potential areas of confusion. This information can be critical to mitigate harms such as undecipherable controls, not meeting user privacy expectations, or worse, abusing user privacy through confusing, obscure structures. 

\subsubsection{Using \tool{} to inform large-scale survey design} Our findings uncovered important aspects that can be used to inform larger-scale surveys to characterize digital norms or expectations (e.g., \cite{apthorpe2018discovering, abdi2021privacy, kuhtreiber2025multi}). For example, we found that participants did not see Amazon as an information technology company managing their information exchanges with the VFAI. This information is critical for designing a survey about VFAIs, otherwise it can lead to misunderstandings. These misunderstandings may not surface in standard interviews without prompts such as our cards. We were also able to find specific pieces of content (attributes) that are likely to matter to older adults in ways that those born with pocket computers may not think twice about, such as timestamps. Qualitative, one-on-one sessions with participants using protocols similar to \tool~ may thus be a possible path towards creating more user-friendly surveys that capture more in-tune data.

\subsubsection{Using \tool~ to include user groups that have been historically overlooked and marginalized in the design of technology} Our participants were older adults who had just recently become familiarized with VFAIs through our study. Even though these devices hold great promise for older adults, their design have overlooked older adults~\cite{pradhan2018accessibility}. Our cards were tailored to our older adult participants, we used words they used in our previous interviews (e.g., ``calling'' the VFAI), and personalized the cards based on their specific usage logs (e.g., if participant had been using Spotify, we included Spotify as a \textit{recipient}). In addition, approaches similar to \tool~ could be used with other user groups that have been underrepresented and marginalized in the design process. 


\subsubsection{Using \tool~ to operationalize the contextual integrity theory of privacy for use by non-technical institutions} Protocols based on \tool~ could be used by enterprises and institutions that are increasingly integrating ubiquitous technologies such as VFAIs in service of specific populations (e.g., health care, education, social services) to both improve digital literacy, and design consent and data sharing mechanisms. For example, a senior center director could utilize \tool~ to learn about possible VFAI information flows and what senior center members expect regarding them. \tool~ could also facilitate the development of the necessary language and qualitative data for others in public or protected services to advocate for policies towards more CI-preserving technologies. For example, in our study, using \tool~ helped us identify the necessary language for clarifying that the VFAI (Alexa) is part of a big company (Amazon), and that an employee of that company can be a computer scientist (not necessarily a factory worker or delivery driver). Finding the right language can help unify people towards mutual advocacy for honoring the privacy and integrity of all.

\subsubsection{Using \tool{} to uncover users' mental models and contextual overlap}
Many modern-day ubiquitous technologies operate in many contexts at once, making it more difficult for people, especially those not in computing fields, to conceptualize CI.
For example, if a user, such as P5, were to share an Amazon password with her family members, which she associates with buying things online and dissociates with the VFAI, then the family would also have access to her ``private'' interactions with the VFAI. This would expose her to their judgment, which she tries to avoid. 
CI theory requires that actors be defined by their contextual roles and capacities. Explicitly stating these during conversations with participants revealed confusion regarding the role of the VFAI. Without addressing this confusion, and clarifying the ties between products like VFAIs and the companies that make them, it is difficult to obtain an accurate signal about users' privacy expectations. 

\section{Limitations}
We conducted this study in an urban part of the U.S. with a small number of participants. Although we had a high level of familiarization with our participants and their VFAI perceptions and usage patterns, more research is needed to determine how they generalize to other user groups around the world, especially given privacy expectations in different cultural settings. Given the small sample size of only older adults, we cannot compare the effectiveness of \tool~ across different age groups. However, given how well our findings aligned with CI, we believe that many of the findings will generalize across age groups and that our design decisions will make \tool~ widely accessible. Similarly, the study was semi-structured, which means that interview questions varied based on what participants responded, and how able they were to stay on track and make judgements. We did not go through as many questions with every participant as we originally set out to, because some interviews took much longer than anticipated. Despite these limitations, our findings shed light on critical issues regarding designing ethical VFAIs. 
\section{Conclusion}
We created \tool~ to explore privacy concerns of VFAIs to support aging in place. We describe \tool' iterative design process, which explains how we adapted them to older adults based on their usage logs. In designing \tool~ for older adults, we
made highly technical CI concepts accessible to them through interactive scenarios grounded in participants' lived experiences.
We uncovered several ethical concerns of using VFAIs for older adults, such as insufficient mental models that affect their ability to properly consent. We also identified implications for design and research, including using \tool~ to inform large-scale survey design.
Our novel empirical findings and their associated implications for the research and design of VFAIs provide evidence of the effectiveness of \tool. Our work helps characterize the dynamic nature of privacy perceptions, the impact of emotional attachment on privacy decisions, and the need to more accurately represent how VFAIs work. \tool{} and our field study findings advance inclusive design practices, providing practical and timely information to achieve ethical VFAIs that support aging in place.

\begin{acks}
We sincerely thank our study participants, who made this work possible, and the senior center directors who helped us connect with them. We also thank the reviewers for their thoughtful feedback, which strengthened this work. We gratefully acknowledge our colleagues for their contributions to data analysis (Jessie Taft), device deployment (Hyein Baek), study design (Nicki Dell), CI theory of privacy guidance (Helen Nissenbaum), conceptual framing (Lynn Andrea Stein), and publication strategy (James Landay). In addition, we thank Sarah Bloomer and Cheryl Spector for their support with proofreading and editing. Yan Shvartzshnaider is supported by the Natural Sciences and Engineering Research Council of Canada (NSERC) grant, RGPIN-2022-04595. Deborah Estrin was supported by the multi-site CREATE Center under the National Institute on Aging, Award Number 3P01AG073090-02S1. Andrea Cuadra and Samar Sabie were both Digital Life Initiative (DLI) Doctoral Fellows when this project was being developed, and would like to thank the DLI group for their valuable feedback on the design of \tool. 
\end{acks}

\bibliographystyle{ACM-Reference-Format}
\bibliography{bibfile}

\end{document}